\documentclass[preprint2]{aastex6}

\usepackage{color}
\usepackage{graphicx}
\usepackage{booktabs}

\usepackage[normalem]{ulem}

\definecolor{red}{rgb}{1.0,0.0,0.0}

\turnoffedit

\shorttitle{The dust ring around HD~157587}
\shortauthors{Millar-Blanchaer et al.}

\begin{document}

\title{Imaging an 80 AU radius dust ring around the F5V star HD~157587}

\author{Maxwell A. Millar-Blanchaer\altaffilmark{1,2}, Jason Wang\altaffilmark{3}, Paul Kalas\altaffilmark{3,4}, James R. Graham\altaffilmark{3}, Gaspard Duch{\^e}ne\altaffilmark{3,5}, Eric L. Nielsen\altaffilmark{4,6}, Marshall Perrin\altaffilmark{7}, Dae-Sik Moon\altaffilmark{1}, Deborah Padgett\altaffilmark{8}, Stanimir Metchev\altaffilmark{9,10}, S. Mark Ammons\altaffilmark{11}, Vanessa P. Bailey\altaffilmark{6}, Travis Barman\altaffilmark{12}, Sebastian Bruzzone\altaffilmark{9}, Joanna Bulger\altaffilmark{13}, Christine H. Chen\altaffilmark{7}, Jeffrey Chilcote\altaffilmark{2}, Tara Cotten\altaffilmark{14}, Robert J. De Rosa\altaffilmark{3}, Rene Doyon\altaffilmark{15}, Zachary H. Draper\altaffilmark{16,17}, Thomas M. Esposito\altaffilmark{3}, Michael P. Fitzgerald\altaffilmark{18}, Katherine B. Follette\altaffilmark{6}, Benjamin L. Gerard\altaffilmark{16,17}, Alexandra Z. Greenbaum\altaffilmark{19}, Pascale Hibon\altaffilmark{20}, Sasha Hinkley\altaffilmark{21}, Li-Wei Hung\altaffilmark{18}, Patrick Ingraham\altaffilmark{22}, Mara Johnson-Groh\altaffilmark{16}, Quinn Konopacky\altaffilmark{23}, James E. Larkin\altaffilmark{18}, Bruce Macintosh\altaffilmark{6}, J{\'e}r{\^o}me Maire\altaffilmark{2}, Franck Marchis\altaffilmark{4}, Mark S. Marley\altaffilmark{24}, Christian Marois\altaffilmark{17,16}, Brenda C. Matthews\altaffilmark{17,16}, Rebecca Oppenheimer\altaffilmark{25}, David Palmer\altaffilmark{11}, Jennifer Patience\altaffilmark{26}, Lisa Poyneer\altaffilmark{11}, Laurent Pueyo\altaffilmark{7}, Abhijith Rajan\altaffilmark{26}, Julien Rameau\altaffilmark{15}, Fredrik T. Rantakyr{\"o}\altaffilmark{27}, Dmitry Savransky\altaffilmark{28}, Adam C. Schneider\altaffilmark{29}, Anand Sivaramakrishnan\altaffilmark{7}, Inseok Song\altaffilmark{14}, Remi Soummer\altaffilmark{7}, Sandrine Thomas\altaffilmark{22}, David Vega\altaffilmark{4}, J. Kent Wallace\altaffilmark{30}, Kimberly Ward-Duong\altaffilmark{26}, Sloane Wiktorowicz\altaffilmark{31}, and Schuyler Wolff\altaffilmark{19}}

\altaffiltext{1}{Department of Astronomy \& Astrophysics, University of Toronto, Toronto, ON M5S 3H4, Canada}
\altaffiltext{2}{Dunlap Institute for Astronomy \& Astrophysics, University of Toronto, Toronto, ON M5S 3H4, Canada}
\altaffiltext{3}{Astronomy Department, University of California, Berkeley; Berkeley CA 94720, USA}
\altaffiltext{4}{SETI Institute, Carl Sagan Center, 189 Bernardo Avenue,  Mountain View, CA 94043, USA}
\altaffiltext{5}{Univ. Grenoble Alpes/CNRS, IPAG, F-38000 Grenoble, France}
\altaffiltext{6}{Kavli Institute for Particle Astrophysics and Cosmology, Department of Physics, Stanford University, Stanford, CA, 94305, USA}
\altaffiltext{7}{Space Telescope Science Institute, Baltimore, MD 21218, USA}
\altaffiltext{8}{Goddard Space Flight Center, 8800 Greenbelt Rd, Greenbelt, MD 20771, USA}
\altaffiltext{9}{Department of Physics and Astronomy, Centre for Planetary Science and Exploration, the University of Western Ontario, London, ON N6A 3K7, Canada}
\altaffiltext{10}{Department of Physics and Astronomy, Stony Brook University, Stony Brook, NY 11794-3800, USA}
\altaffiltext{11}{Lawrence Livermore National Laboratory, Livermore, CA 94551, USA}
\altaffiltext{12}{Lunar and Planetary Laboratory, University of Arizona, Tucson AZ 85721, USA}
\altaffiltext{13}{Subaru Telescope, NAOJ, 650 North A'ohoku Place, Hilo, HI 96720, USA}
\altaffiltext{14}{Department of Physics and Astronomy, University of Georgia, Athens, GA 30602, USA}
\altaffiltext{15}{Institut de Recherche sur les Exoplan{\`e}tes, D{\'e}partment de Physique, Universit{\'e} de Montr{\'e}al, Montr{\'e}al QC H3C 3J7, Canada}
\altaffiltext{16}{University of Victoria, 3800 Finnerty Rd, Victoria, BC V8P 5C2, Canada}
\altaffiltext{17}{National Research Council of Canada Herzberg, 5071 West Saanich Rd, Victoria, BC V9E 2E7, Canada}
\altaffiltext{18}{Department of Physics \& Astronomy, University of California, Los Angeles, CA 90095, USA}
\altaffiltext{19}{Department of Physics and Astronomy, Johns Hopkins University, Baltimore, MD 21218, USA}
\altaffiltext{20}{European Southern Observatory , Alonso de Cordova 3107, Vitacura, Santiago, Chile}
\altaffiltext{21}{School of Physics, College of Engineering, Mathematics and Physical Sciences, University of Exeter, Exeter EX4 4QL, UK}
\altaffiltext{22}{Large Synoptic Survey Telescope, 950N Cherry Av, Tucson, AZ 85719, USA}
\altaffiltext{23}{Center for Astrophysics and Space Science, University of California San Diego, La Jolla, CA 92093, USA}
\altaffiltext{24}{Space Science Division, NASA Ames Research Center, Mail Stop 245-3, Moffett Field CA 94035, USA }
\altaffiltext{25}{American Museum of Natural History, Depratment of Astrophysics, New York, NY 10024, USA}
\altaffiltext{26}{School of Earth and Space Exploration, Arizona State University, PO Box 871404, Tempe, AZ 85287, USA}
\altaffiltext{27}{Gemini Observatory, Casilla 603, La Serena, Chile}
\altaffiltext{28}{Sibley School of Mechanical and Aerospace Engineering, Cornell University, Ithaca, NY 14853, USA}
\altaffiltext{29}{Department of Physics and Astronomy, University of Toledo, Toledo, OH 43606, USA}
\altaffiltext{30}{Jet Propulsion Laboratory, California Institute of Technology Pasadena CA 91125, USA}
\altaffiltext{31}{The Aerospace Corporation, 2310 E. El Segundo Blvd., El Segundo, CA 90245}

\email{$^{1}$maxmb@astro.utoronto.ca}

\begin{abstract}

We present $H$-band near-infrared polarimetric imaging observations of the F5V star HD~157587 obtained with the Gemini Planet Imager (GPI) that reveal the debris disk as a bright ring structure at a separation of $\sim$80$-$100~AU. The new GPI data complement recent HST/STIS observations that show the disk extending out to over 500~AU. The GPI image displays a strong asymmetry along the projected minor axis as well as a fainter asymmetry along the projected major axis. We associate the minor and major axis asymmetries with polarized forward scattering and a possible stellocentric offset, respectively. To constrain the disk geometry we fit two separate disk models to the polarized image, each using a different scattering phase function. Both models favor a disk inclination of $\sim 70\degr$ and a $1.5\pm0.6$ AU stellar offset in the plane of the sky along the projected major axis of the disk. We find that the stellar offset in the disk plane, perpendicular to the projected major axis is degenerate with the form of the scattering phase function and remains poorly constrained. The disk is not recovered in total intensity due in part to strong adaptive optics residuals, but we recover three point sources. Considering the system's proximity to the galactic plane and the point sources' positions relative to the disk, we consider it likely that they are background objects and unrelated to the disk's offset from the star. 

\end{abstract}


\section{Introduction}

Circumstellar debris disks, composed of planetesimals and dust, are remnants of the planet formation process. Therefore, their study can provide insights into the planet formation and evolution history of the systems in which they reside. The dust grain composition of a disk traces grain growth and erosion, and, if spatially resolved, disk morphology can provide evidence of dynamical interactions with nearby planets. Such an interaction can manifest as a warp \citep[e.g. Beta Pic;][]{Burrows1995, Mouillet1997}, a stellocentric offset \citep[e.g. HR 4796A;][]{Wyatt1999, Telesco2000}  or a sharp radial profile at the inner edge of a dust ring \citep[e.g. Fomalhaut;][]{Kalas2005,Quillen2006}. 

\begin{figure*}[Ht]
\includegraphics[width=\linewidth]{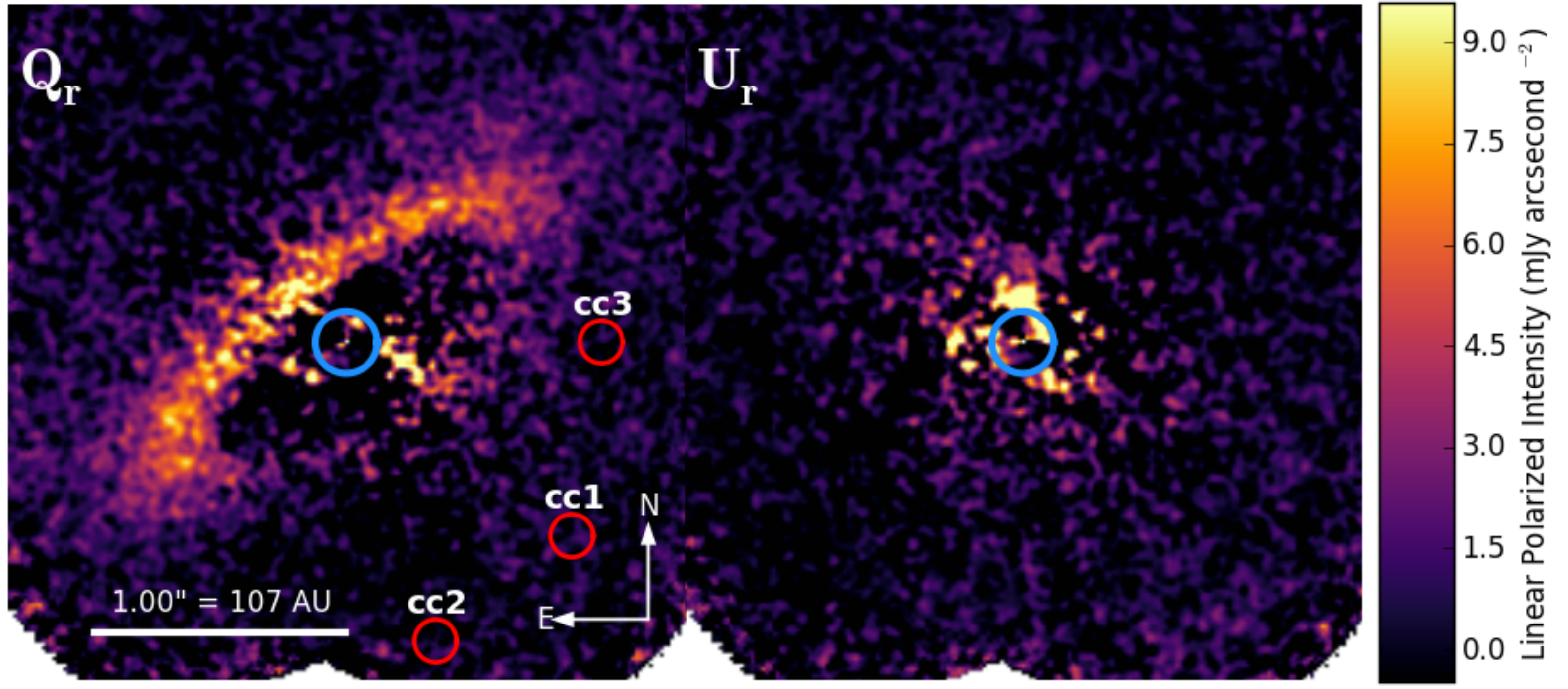}
\caption{\emph{Left}: 
GPI $H$-band radial polarized intensity image ($Q_r$) of the HD~157587 debris disk. The image 
has been smoothed with a Gaussian kernel ($\sigma$ = 1~pixel). The blue circle indicates the size of the central focal plane mask ($0\farcs12$ radius). The red circles denote the point sources seen in total intensity in Figure~\ref{fig:stokesi} (Section~\ref{sec:pyklip}). 
\emph{Right}: The $U_r$ image of HD~157587. For single scattering from circumstellar material we expect no contributions to the $U_r$ image. Thus, this image can be used as a noise map for the $Q_r$ image. The image appears to be largely free of correlated structure, except at small inner working angles. Both images have been cropped from the full GPI field of view to display only the inner $2\farcs6 \times 2\farcs6$ region. No polarized emission was seen outside the cropped region. \label{fig:diskimg}}
\end{figure*}

Debris disks are imaged via their thermal emission in infrared or millimeter wavebands, which typically traces the location of millimeter sized bodies, or via scattered light in the visible and near-infrared (NIR), which is more sensitive to micron-sized dust. Observations of debris disks in scattered light are typically able to resolve finer spatial scales than longer wavelength observations (though ALMA's spatial resolution is now competitive), but are challenging due to the extreme contrast ratios between the faint dust-scattered light and the bright host stars. Instrumental point-spread functions (PSFs) extend the stellar emission out to angular separations where debris disks are found, obscuring the scattered light from the dust. For ground-based observations this problem is compounded by the atmosphere, which scatters light from the PSF out to farther separations. 

The Gemini Planet Imager \citep[GPI;][]{Macintosh2014} is an instrument on the Gemini South 8-m telescope that has been designed specifically to mitigate these challenges. It employs a high-order adaptive optics (AO) system, combined with an apodized-pupil Lyot coronagraph and an integral-field spectrograph, to image exoplanets and debris disks at angular separations down to $\sim0\farcs1$. 
The GPI Exoplanet Survey (GPIES) is a long-term Gemini South program targeting 600 nearby stars with the goal of discovering and characterizing young Jovian exoplanets. A secondary goal of the survey is to image and characterize debris disks. Stars with previously resolved debris disks and survey stars that exhibit infrared excesses are observed using GPI's polarimetry mode. The polarimetry mode is implemented as a rotatable half-wave plate (HWP) modulator and a Wollaston prism analyser. This mode has been designed to take advantage of the inherent polarization of light scattered off circumstellar dust grains, to further suppress the unpolarized starlight and reveal the disk beneath \citep{Perrin2015}.  

Here we present GPIES observations of the debris disk around HD~157587, an F5V star with an infrared excess $L_{IR}/L_{star}$= 7.9$\times10^{-4}$, \citep{mcdonald12a} at a distance of 107.4~pc \citep{vanLeeuwen2007}. HST/STIS coronagraphic imaging (GO-12998; PI Padgett) first revealed the dust scattered light extending to $>$7$''$ radius, with a morphology resembling a fan \citep[such as for HD~15745;][]{kalas07a}, where the straight edge of the fan lies along the southwestern side of the nebulosity \citep{Padgett2015}. The inner working angle of these data corresponds to a projected separation of $\sim$100 AU. Our new scattered light images, obtained as part of the GPIES campaign, detect the structure of the circumstellar dust in the projected 30 - 130 AU radial region. 


\section{Observations and Data Reduction}
\label{sec:obsdata}
We observed HD~157587 with GPI's polarimetry mode in the $H$-band on 2015 August 28 UT. The observations consisted of twenty-eight 90~s frames, with the HWP position angle cycling between $0\fdg0$, $22\fdg5$, $45\fdg0$ and $67\fdg5$. Throughout the sequence the field rotated by a total of $46\degr$. The average airmass was 1.02 and the seeing as measured by the Gemini Differential Image Motion Monitor and Multi-Aperture Scintillation Sensor was $0.61\arcsec$ and $0.63\arcsec$, respectively. The AO system telemetry reported a post-correction wavefront rms error of $216\pm20$ nm across the sequence. 

The data were reduced using the GPI data reduction pipeline version 1.3 \citep{MaireSPIE2012, PerrinSPIE2014}. The raw data were dark subtracted, cleaned of correlated detector noise, bad pixel corrected, flexure corrected and then combined into a polarization datacube (where the third dimension holds two orthogonal polarization states). Each datacube was divided by a polarized flat field and corrected for non-common path errors via a double differencing algorithm \citep{Perrin2015}. The instrumental polarization was determined by estimating the apparent stellar polarization in each polarization datacube by measuring the mean normalized difference of pixels with separations between 7 and 13 pixels from the star's location (determined from the satellite spots using a radon-transform-based algorithm; \citealt{Wang2014}). The estimated instrumental polarization was then subtracted from each pixel, scaled by the pixel's total intensity \citep{MillarBlanchaer2015}. The region selected to measure the instrumental polarization was just outside of the coronagraph edge where the residual PSF flux, and hence the flux from instrumental polarization, is maximized. We assume that this area is devoid of any significant polarized structure and that any measured difference between the two polarization states is due to the instrumental polarization. 

The datacubes were corrected for geometric distortion, smoothed with a Gaussian kernel ($\sigma = 1$~pixel) and then combined into a Stokes datacube by solving a set of equations that describe the linear polarization states measured in each of the individual exposures given the waveplate and sky rotation angles \citep{Perrin2015}. The Stokes datacube was subsequently converted to the radial Stokes convention \citep[$\lbrack I,Q,U,V\rbrack \rightarrow \lbrack I,Q_r,U_r,V\rbrack$;][]{Schmid2006}. The sign convention is such that a positive $Q_r$ corresponds to a polarized intensity whose vectors are oriented perpendicular to a line connecting a given pixel to the central star and negative values are parallel to the line. Under this convention (and the assumption of low optical depth) the $U_r$ image should contain no disk flux and will only contain noise. Thus the $Q_r$ image should contain all of the disk polarized intensity as positive values. Finally, the flux of the four satellite spots was measured and flux calibration was carried out as described in \citet{Hung2015}. The final $Q_r$ and $U_r$ images can be seen in Figure~\ref{fig:diskimg}. 

The polarization datacubes were also processed separately using the \texttt{pyKLIP} \citep{pyKLIP} implementation of the Karhunen-Lo\`eve Image Projection (KLIP) algorithm \citep{Soummer2012} to attempt to recover the disk in total intensity and search for point sources (Section \ref{sec:pyklip}). 


\section{Results}
\subsection{Polarized Intensity Image}

The $Q_r$ image displays an inclined, ring-like structure with a strong brightness asymmetry in the NE-SW direction (the projected minor axis). The inner edge of the ring has projected semi-major and semi-minor axes of $\sim0\farcs65$ and $\sim0\farcs2$, respectively. The region interior to the inner edge of the ring appears to be cleared of any scattering material. However, the residual systematics at smaller separations in the $Q_r$ and $U_r$ images do not exclude additional dust at these smaller radii. Outside of the ring the surface brightness decreases quickly and reaches the noise floor within our field of view (FOV), which extends to a radius of $1\farcs7$ along the semi-major axis of the the ring.


\begin{figure}[t]
\includegraphics[width=\columnwidth]{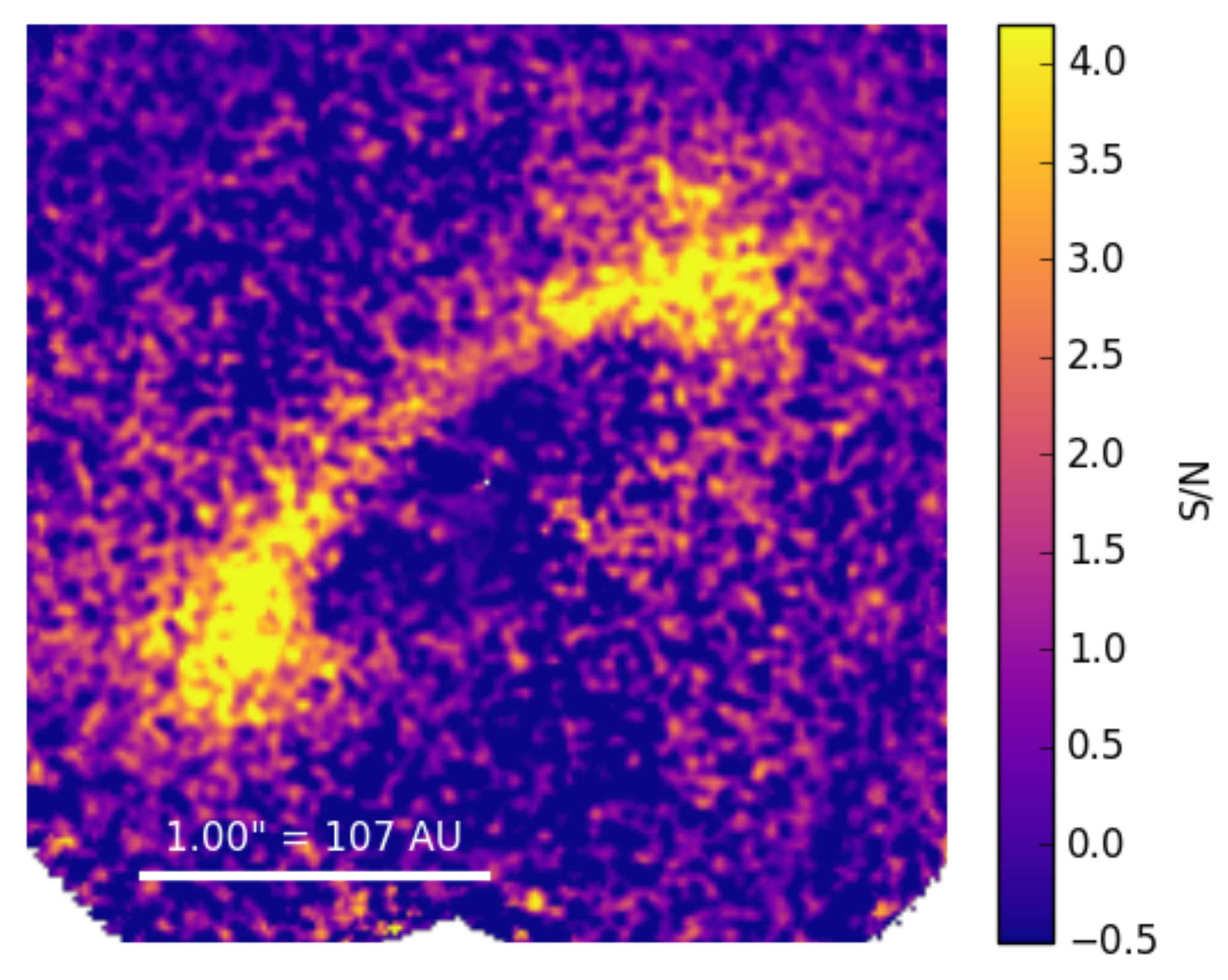}
\caption{Signal to noise map of the 
GPI $H$-band $Q_r$ image. The ansae are clearly detected here with a S/N greater than 5, while the region near the minor axis is closer to a S/N of 3. In this map the projected minor axis (in the NE direction) appears at a low S/N even though it is the brightest region in the $Q_r$ image. This is due to the elevated noise in the inner regions of the $U_r$ image that overlap with the edge of the disk. \label{fig:SNR}}
\end{figure}

A comparison between the $Q_r$ and $U_r$ images indicates that the ring detection is robust and that the morphology is not due to instrumental effects. The $U_r$ image appears to be dominated by uncorrected systematics interior to $\sim0\farcs275$, a region which intersects the ring near its minor axis. Outside of $\sim0\farcs275$, there appear to be no coherent structures in the $U_r$ image.

The strong NE-SW asymmetry seen in the polarized images is reminiscent of the asymmetries seen in other disks recently imaged in polarized light by GPI, for example HR 4796A, \citet{Perrin2015}; HD~106906, \citet{Kalas2015}; HD~131835, \citet{Hung2015}; and HD~61005 (Esposito et al., in press). In all of these disks, this asymmetry is interpreted as the disk being tilted such that the brighter side is closer to the observer and the observed brightness asymmetry is mostly due to strong forward scattering in the polarized scattering phase function. Indeed, a recent analysis of Cassini observations (albeit \emph{total} intensity \emph{visible} light observations) of Saturn's G and D rings indicate that collisionally generated dust is expected to be strongly forward scattering \citep{Hedman2015}. 

In addition to the NE-SW asymmetry, the $Q_r$ image also displays a mild brightness asymmetry between the SE and NW sides of the disk, visible as two main features: a) The SE ansa appears brighter and reveals more of the backside of the disk than the NW ansa, causing the ansa to appear hook-like, and b) the SE side of the disk appears brighter along the bright NE edge of the disk, about the NE semi-minor axis. These features are confirmed in a signal-to-noise ratio (S/N) map (Figure~\ref{fig:SNR}), created by dividing the $Q_r$ image at each point by the standard deviation of an annulus in the $U_r$ image at the same angular separation. 


\begin{figure}[t]
\includegraphics[width=0.9\linewidth]{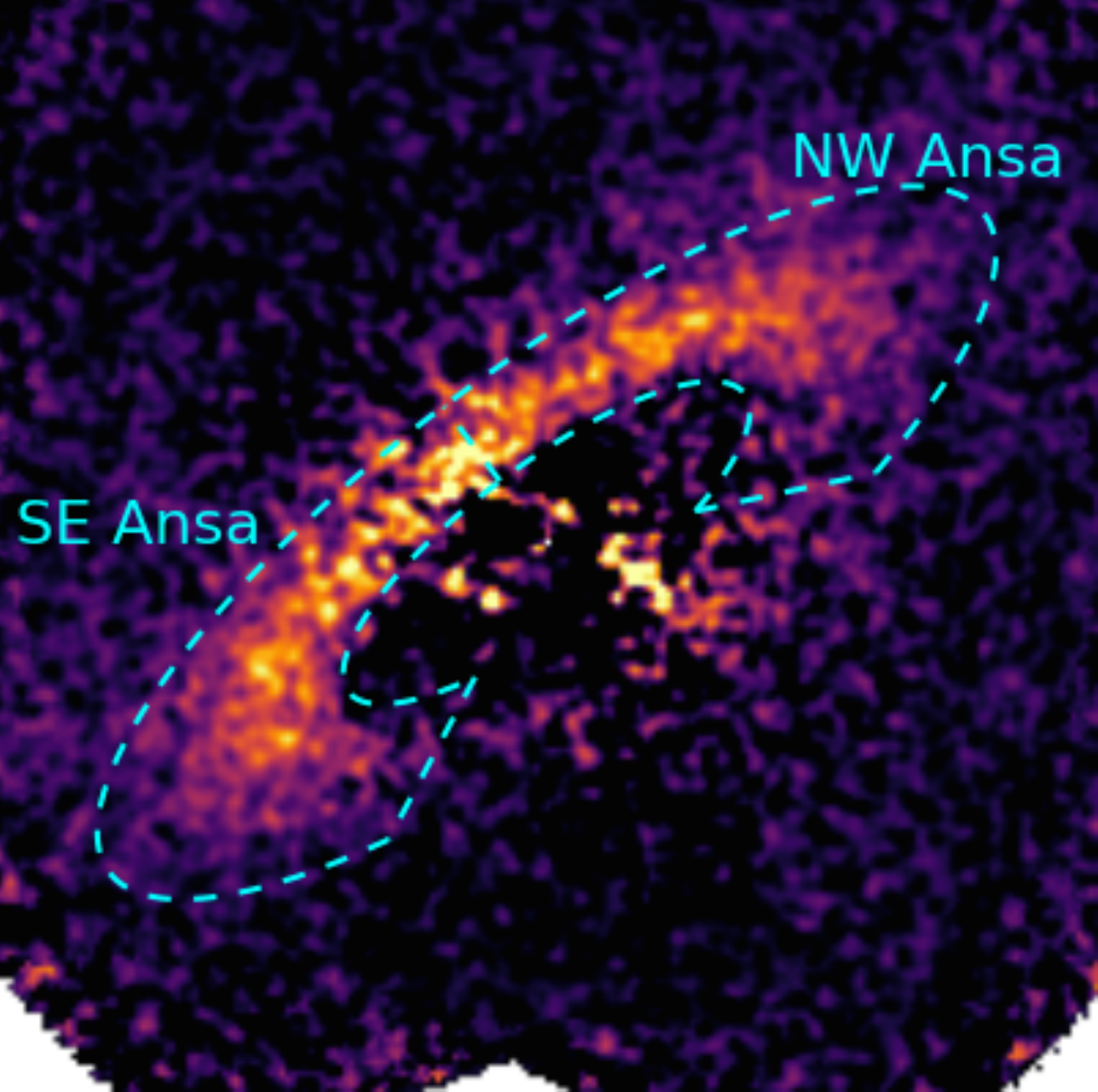}
\caption{The regions used to calculate the magnitude of the brightness asymmetry between the two ansae overplotted on top of the $Q_r$ image from Figure~\ref{fig:diskimg}.
\label{fig:regions}}
\end{figure}

To estimate the magnitude of this asymmetry we created a custom-shaped aperture for each ansa (Figure~\ref{fig:regions}). The two apertures are mirror images of each other, with the axis of symmetry coincident with the projected minor axis (a position angle of 37$\degr$, Section~\ref{sec:modeling}). By summing the flux in each aperture, we find the brightness ratio between the SE ansa and the NW ansa to be $1.15\pm0.02$, where the errors at each pixel are calculated in the same manner as when creating the S/N map. 

This brightness asymmetry may be explained by a stellar offset, which in turn may be caused by a perturbing planet in an eccentric orbit that imparts a forced eccentricity to the dust's parent bodies (e.g. \citealt{Wyatt1999}). For small eccentricities, the morphology of the disk remains axisymmetric (to first order), but the host star is no longer located at the geometric center of the disk. A brightness asymmetry can then be seen as a result of one side of the disk being closer to the star than the other and receiving increased stellar irradiation. This offset also warms the closer dust gains, an effect know as pericenter glow which can be observed in thermal emission \citep{Wyatt1999}.

\subsection{Total Intensity Image}
\label{sec:pyklip}
Each individual polarization datacube was summed across its two polarization channels to create a total intensity datacube. The entire set was then processed with \texttt{pyKLIP} using a large range of Karhunen-–Lo\`{e}ve (KL) modes, exclusion criteria and optimization regions. No disk emission was detected in any reduction. This is due in part to prominent stellar residuals resulting from imperfect AO correction caused by strong winds throughout the observation set. These winds held a roughly constant position angle during the observations, which caused the AO residuals to rotate relative to the instrument frame, mimicking the rotation of an astrophysical source. Aggressive PSF subtraction is able to suppress nearly all of this signal, but also suppresses any of the disk emission. In addition to the wind residuals, any angular differential imaging-based method will be subject to self-subtraction for such an azimuthally extended disk, compounding the difficulties in detecting the disk in total intensity. 

\begin{figure}[t]
\includegraphics[width=\linewidth]{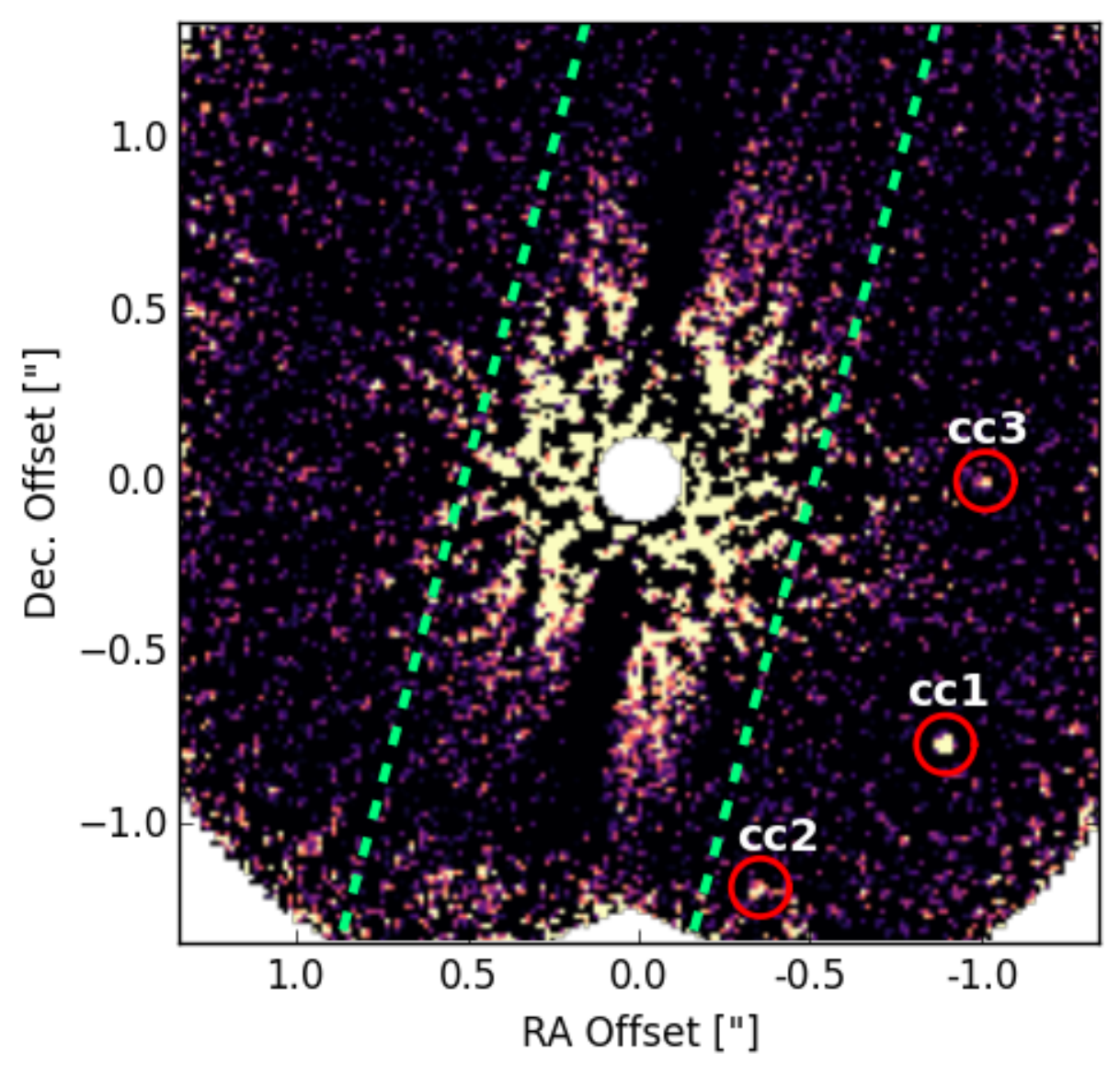}
\caption{The PSF-subtracted total intensity (Stokes $I$) image of HD~157587 at $H$-band. No disk is detected, in part because of strong AO residuals caused by winds. Residuals due to the winds can be seen as nearly vertical dark streaks to the north and south of the obscured star. The three point-sources described in Table~\ref{tab:gois} are marked by the red circles. \edit1{We consider the region inside the green dashed lines to be dominated by wind residuals.} \label{fig:stokesi}}
\end{figure}

Although no dust-scattered light was detected, we recovered three \edit1{possible} point sources in the PSF-subtracted data (Figure~\ref{fig:stokesi}). Their measured properties are summarized in Table~\ref{tab:gois}\edit1{, where we have labeled them candidate companions (cc) 1, 2 and 3}. \edit1{The brightest of the sources, cc1, is confidently detected. Due to its position on the edge of the field of view, cc2 is just below a $5\sigma$ detection.} The faintest of the three sources (source \edit1{cc3}) is recovered at less than $3\sigma$ significance. However, we find that \edit1{both cc2 and cc3 are} stable as a function of KL modes and \edit1{appear as point sources} in both our most conservative (i.e., with low number of KL modes) and aggressive (i.e., with a high number of KL modes) reductions, which does not hold true for other low significance point source candidates in the data. \edit1{Thus, we have decided to report both alongside the one confident detection.}

\begin{deluxetable*}{lccccc}
\tablecaption{Candidate Point Source Properties \label{tab:gois}} 
\tablehead{
\colhead{Label} & \colhead{S/N} & \colhead{Separation} & \colhead{Position Angle} & \colhead{$H$-band Flux Ratio\tablenotemark{a}} & \colhead{Radial Separation\tablenotemark{b}} \\
}
\startdata
cc1 & 6.6  & $1\farcs180 \pm 0\farcs002 $ & $228\fdg9 \pm 0\fdg2$ & $(3.2\pm0.8)\times 10^{-6}$ & 364 AU \\
cc2 & 4.2 & $1\farcs248 \pm 0\farcs005 $ &  $195\fdg9 \pm 0\fdg2$ & $(2.7\pm0.7)\times 10^{-6}$ & 380 AU \\
cc3 & 2.8  & $1\farcs002 \pm 0\farcs004 $ & $269\fdg3 \pm 0\fdg3$ & $(1.9\pm0.7)\times 10^{-6}$ & 210 AU\\
\enddata
\tablenotetext{a}{Between the source and the star} 
\tablenotetext{b}{If on a circular orbit in the disk plane} 
\end{deluxetable*}

The flux and position of the point sources were calculated using a Gaussian matched filter. The S/N was determined by comparing the flux of the point sources with the noise at the same radial separation. Because the point sources lie outside of the region with strong wind residuals, we used a parallelogram-shaped region to mask out the wind residuals when estimating the noise \edit1{(see Figure \ref{fig:stokesi})}. To correct for algorithm throughput and to characterize the uncertainties, artificial point sources of known brightness and position were injected into the data at similar separations but at different azimuthal positions with respect to the point sources \edit1{cc1, cc2 and cc3}, avoiding the region with strong wind residuals. Algorithm throughput was estimated by measuring the flux of the artificial point sources after PSF subtraction. The scatter in the position and flux of the artificial planets were used as the uncertainties on the position and flux of the point sources, respectively. To obtain the total error in the astrometry, we used the reported plate scale and North angle from \citet{derosa2015} and added the uncertainties in quadrature. For our flux conversion, we used the flux of the satellite spots to convert the flux of the point sources to contrast units, using the standard GPI calibrations for the flux ratio of the satellite stars relative to the central PSF \citep{Wang2014}. The scatter in the satellite spot fluxes was used as the uncertainty in the flux conversion factor. 

While it is possible that one or more of these three point sources is associated with HD~157587, we note that the star's projected position on the sky is near the galactic plane ($[l,b] = [6\fdg0, 9\fdg4]$) and it is likely that most, if not all, of these sources are background objects. We further discuss the potential relationship of these point sources to the debris disk in Section~\ref{sec:discussion}.

\begin{figure}[t]
\includegraphics[width=\linewidth]{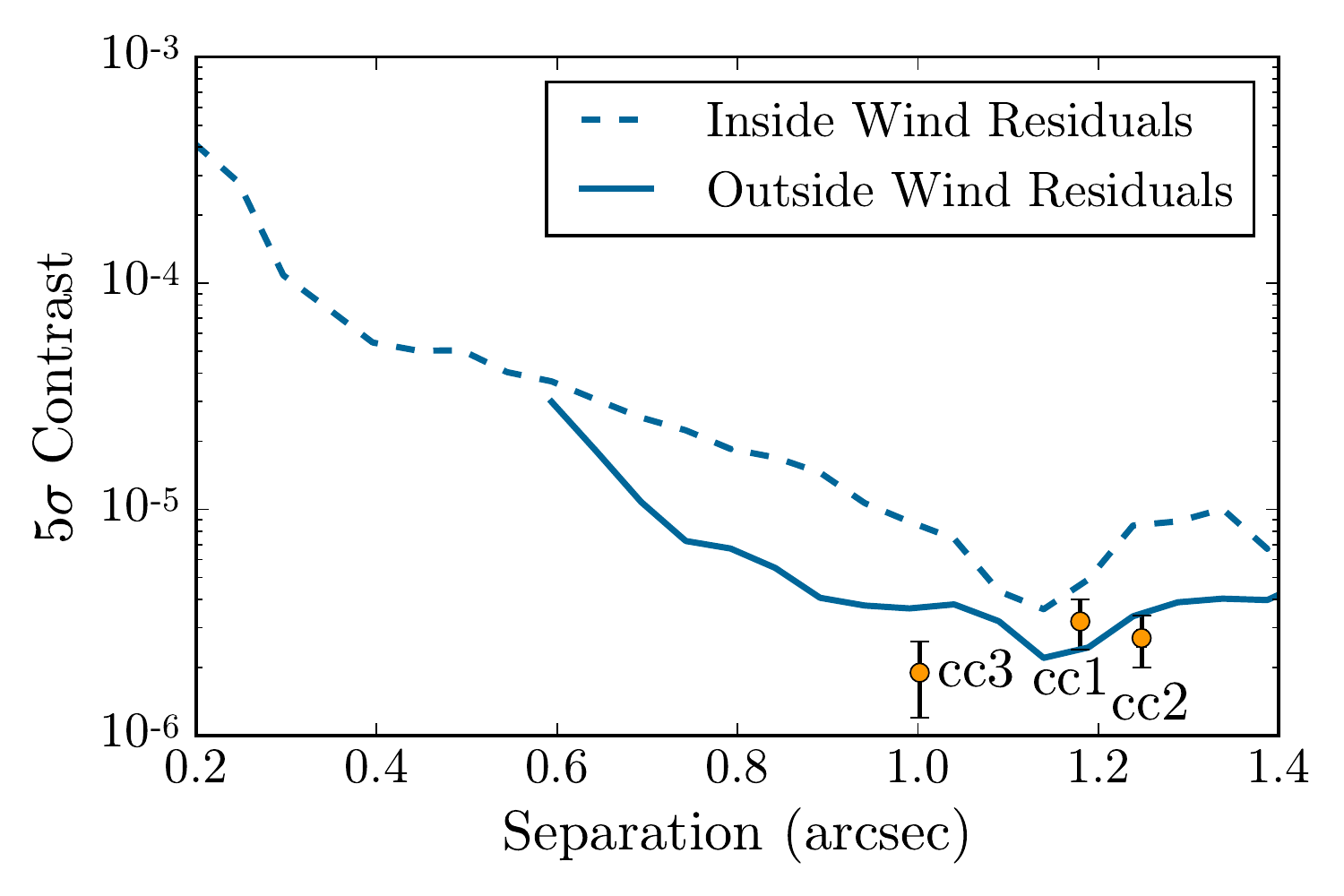}
\caption{\edit1{Sensitivity to point sources in the total intensity image. The dashed and solid blue lines are the sensitivity curves inside and outside of the wind residuals, respectively. The candidate point sources are plotted in orange. Consistent with their reported S/N in Table \ref{tab:gois}, only b detected above a significance of $5\sigma$.} \label{fig:contrast}}
\end{figure}

\edit1{To assess our sensitivity to additional point sources in the image, we generated contrast curves for our total intensity image (Figure \ref{fig:contrast}). Because of the strong wind residuals, we computed contrast curves both inside and outside of the wind residuals using the aforementioned parallelogram shape (seen in Figure~\ref{fig:stokesi}) to define the two regions. Injected point sources were again used to correct for algorithm throughput and the reported $5\sigma$ contrasts were corrected for small number statistics \citep{Mawet2014}. }

\section{Disk Modeling}
\label{sec:modeling}
To recover basic geometric properties of the disk, we modeled the $Q_r$ image using two modified versions of the disk model presented in \citet{MillarBlanchaer2015}. The original model describes the three-dimensional dust density as a radial power law centered on the host star with a Gaussian height profile and constant aspect ratio (the ratio of the disk scale height to the radial separation). Optically thin (single) scattering is assumed and a Henyey-Greenstein (HG) function is used as a polarized scattering phase function. A disk image is calculated by combining the dust density profile with the scattering phase function and integrating along the line of sight. The model includes nine free parameters: inner radius, $R_1$; outer radius, $R_2$; power law index for the radial dust distribution, $\beta$; disk aspect ratio, $h_0$; average scattering cosine, $g$; inclination, $\phi$; position angle on the sky, $\theta_{PA}$; a flux normalization factor $N_0$ and a constant offset term applied to the entire image $I_0$.

\begin{figure*}[t]
\includegraphics[width=\linewidth]{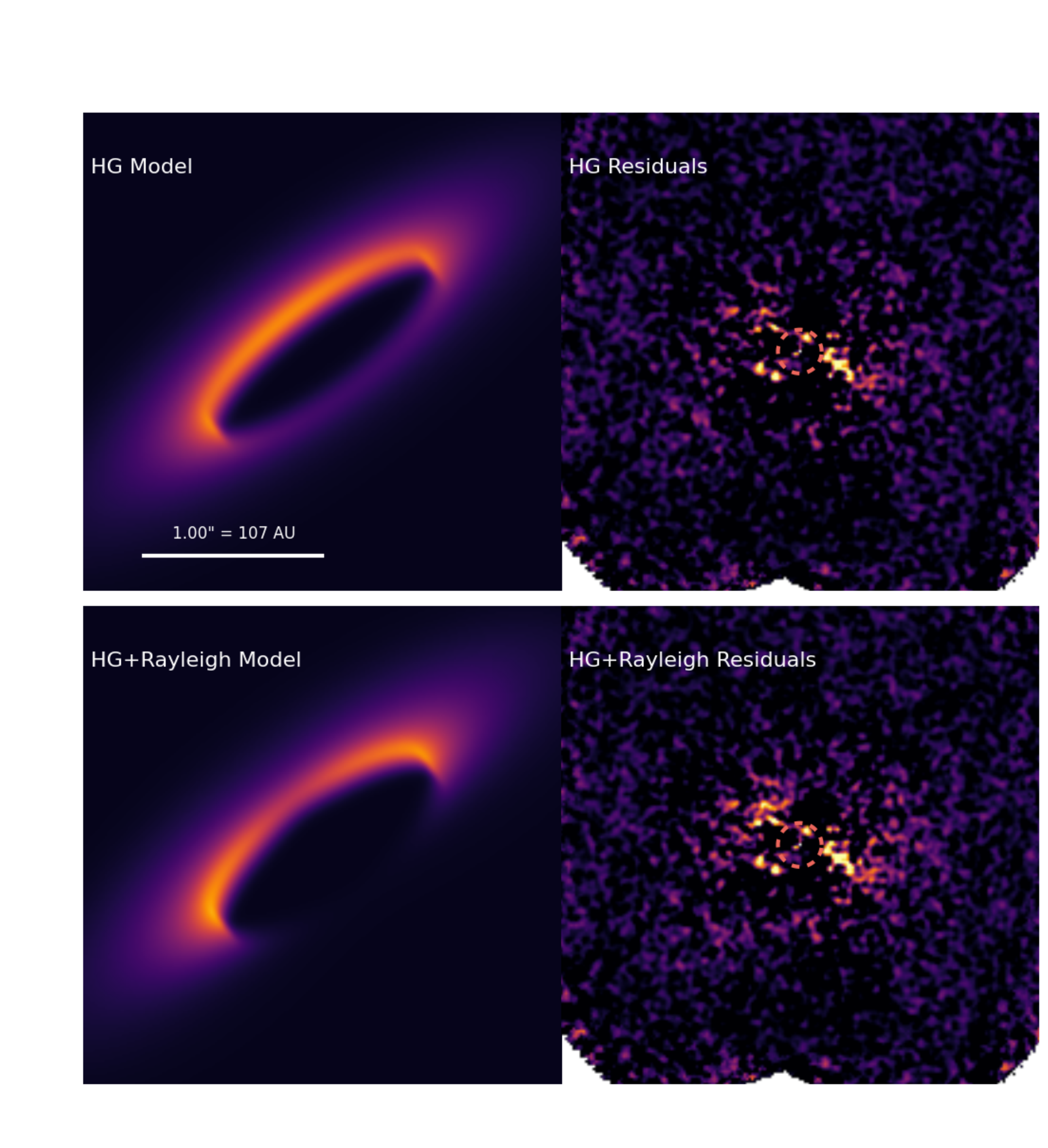}
\caption{\emph{Left}: Images of the best-fit $H$-band, $Q_r$ models described in Table~\ref{tab:model_fits}. \emph{Right}: The residuals of each model subtracted off the $Q_r$ image. The top row displays the disk image and residuals for model which uses only an HG function to describe the polarized scattering phase function. The bottom row displays the disk image and residuals for the model that uses an HG function combined with Rayleigh scattering. The images are displayed at the same colour scale and orientation as Figure~\ref{fig:diskimg}. \label{fig:resids}}
\end{figure*}

Motivated by the SE-NW asymmetries seen in Figure~\ref{fig:diskimg}, we adapted the disk model to allow for a stellar offset in the plane of the disk. This was implemented as two new free parameters: $\Delta X_1$, an offset along the projected semi-major axis of the disk (i.e. parallel to the position angle in the plane of the sky; positive offset is in the SE direction), and $\Delta X_2$, an offset in the disk plane, perpendicular to $\Delta X_1$ (positive offset is towards the backside of the disk). The true orientation of $\Delta X_2$ relative to the observer is parallel to the inclination vector $\phi$ and will have components both along the projected semi-minor axis of the disk and along the observer's line of sight. For example, if the disk is edge-on ($\phi = 90 \degr$), then $\Delta X_2$ is parallel to the line of sight and if the disk is face-on, then $\Delta X_2$ will be in the plane of the sky. 

We considered two different options for the polarized scattering phase function. We first used the HG function, as in the original model,  and we also considered a model where the polarized scattering phase function is described by a Rayleigh scattering function multiplied by an HG function. This second model was considered because the Rayleigh scattering function produces a peak in the polarization fraction at $90\degr$ scattering angles and therefore places increased importance on the ansae, where the hook-feature is seen. In this second model the HG function describes the scattering phase function of the (undetected) total intensity. 

A similar strategy was used by \citet{Graham2007}, who simultaneously fit total intensity and polarization fraction observations of the AU Mic debris disk with the same combination of an HG function and Rayleigh scattering function. One difference between their analysis and ours is that their fit includes the maximum polarization fraction, $p_{max}$, as a free parameter. In our case, with no total intensity detection, we fit only the polarized intensity, not the polarization fraction, and the maximum polarization fraction is folded into the flux normalization term, $N_0$.

Following \citet{MillarBlanchaer2015}, we fit the two disk models to the image using the affine invariant Markov Chain Monte Carlo (MCMC) ensemble sampling package \texttt{emcee} \citep{Foreman-Mackey2013}. The package uses an ensemble of ``walkers" each with its own MCMC chain, where each step of each walker depends on the state of the other walkers. Before fitting, we first applied $3\times3$ pixel binning to both the $Q_r$ image and the $U_r$ image. At $H$-band, the diffraction limit of Gemini is about 3 GPI pixels. Therefore, by binning the data, we can improve S/N without any significant loss of spatial information. 
\edit1{This has the added benefit that each binned pixel can be considered nearly independent, thus correlated errors between the binned pixels are significantly reduced relative to the full-resolution pixels.}
The error estimates for each pixel were calculated as in Section~\ref{sec:obsdata}, by taking the standard deviation of concentric annuli in the (binned) $U_r$ image. Preliminary fitting runs revealed that there was no evidence for an offset term, $I_0$ and as a result we opted to drop the offset as a free parameter. 

For the first model (HG-only), the fitting procedure was run with 240 walkers and a burn-in stage of 500 iterations followed by a full run of 1300 iterations. The burn-in serves to initialize the state of the walkers and only walker positions from the full run are used to calculate the posterior. The fitting code was executed on the Edison supercomputer at the National Energy Research Scientific Computing Center (NERSC), employing one process for each walker. For each iteration, a new model was generated at the same resolution as a full GPI frame, smoothed with a Gaussian filter ($\sigma = 1$~pixel) and then binned $3\times3$, to replicate the final steps of the data reduction. 

After the run, the maximum autocorrelation across all parameters was found to be 65 iterations, indicating that the chains had iterated longer than the required $O(10\times t_{autocorrelation}$) for convergence \citep[see][]{Foreman-Mackey2013}. As a second check,  the ensemble chains were examined by eye after the burn-in stage and appeared to have converged. The entire procedure took 46 hours and 30 minutes to complete using 240 Intel Ivy-Bridge 2.4-GHz cores (10 nodes each with 24 cores). 

We fit the data to the second model (HG + Rayleigh), using the same \texttt{emcee} setup, but only 1000 iterations were completed due to time constraints on the NERSC Edison supercomputer. For this run the autocorrelation time was found to be 91, again, indicating that the chains should have converged. 

\begin{deluxetable*}{lccccc}

\tablecaption{Disk Model Parameters \label{tab:model_fits}} 
\tablehead{
&&&& \multicolumn{2}{c}{Best-Fit Values} \\ \cline{5-6} \\
\colhead{Parameter} & \colhead{Symbol} & \colhead{Units} & \colhead{Prior Range} & \colhead{HG Model} & \colhead{HG + Rayleigh Model}}
\startdata
Inner radius & $R_1$ & $AU$ &$40 - 120$ & $81\pm{2}$ & $78.9\pm{0.8}$ \\
Outer radius & $R_2$ & $AU$ &$R_1 - 1000$ &$216\pm16$ & $211^{+21}_{-15}$\\
Density power law index & $\beta$ & ... & $0.1 - 4$ & $2.23\pm 0.15$ & $2.2\pm0.2$\\
Scale height aspect ratio& $h_0$ & ... &$0.0001 - 0.5$ &$0.079\pm 0.005$& $0.084\pm0.006 $\\
HG asymmetry parameter & $g$ & ... & $0 - 1$ &$0.285\pm0.012$ & $0.65\pm0.03$\\
Line of sight inclination & $\phi$ & $\degr$ & $45 - 85$ & $72.2\pm0.4 $& $68.3^{+0.7}_{-0.8}$\\
Position angle & $\theta_{PA}$ & $\degr$ & $105 - 145\degr$ &$127.0\pm0.3$ & $127.1\pm0.3$\\
Sky plane offset & $\Delta X_1$ & $AU$ & -$R_1 - R_1$ & $ 1.6\pm0.6$ & $1.4\pm0.6$ \\
Line of sight offset & $\Delta X_2$ & $AU$  & -$R_1 - R_1$ & $2.1\pm 1.6$ & $-5.7^{+2.1}_{-2.2}$\\
\midrule
Chi-Squared & $\chi^2$ & ... &  ... & 2648 & 2670 \\
Reduced Chi-Squared & $\chi^2_{red} $ & ... & ... & 0.89 & 0.90
\enddata
\end{deluxetable*}

Marginalized posterior distributions for each parameter of each of the two models  were obtained by sampling the MCMC chains of each walker at intervals equal to one autocorrelation time. A summary of the results of the fitting procedure can be found in Table~\ref{tab:model_fits}, which reports the best-fit value for each parameter as the median of its \edit1{marginalized} posterior, and the errors are taken to be the 68\% confidence intervals. The normalization factor $N_0$ is considered a nuisance parameter and is not included in the table. \edit1{Plots of the marginalized and joint posterior distributions derived from the MCMC chains can be seen in Appendix Figure~\ref{fig:diskfit_pdfs_HG} and Appendix Figure~\ref{fig:diskfit_pdfs_HGRayleigh}, for the HG only and HG + Rayleigh models, respectively. The posteriors for each parameter are all single-peaked and approximately Gaussian, with the exception of the outer radius, $R_2$ that displays a slightly elongated tail towards higher values. In the HG model slight degeneracies are found between $\phi$ and $\Delta X_1$, and $\phi$ and $\Delta X_2$. The HG + Rayleigh model displays addition small degeneracies between $g$ and $\phi$, as well as $g$ and $\Delta X_2$. However, all parameters appear to be well constrained in the marginalized distributions.} Images of the best-fit models for both the HG fit and the HG + Rayleigh fit can be seen in Figure~\ref{fig:resids}, alongside an image of the residuals of each model subtracted from the $Q_r$ image. 

\edit1{Table~\ref{tab:resids_stats} displays the root mean square (RMS) pixel values, as well as the fifth and ninety-fifth percentile pixel values (as a proxy for dynamic range) for the $Q_r$ image, the $U_r$ image, and the two residuals images. The measurements have been split into an inner region (inside 0\farcs275) and outer region (between 0\farcs3 and 1\farcs3). The inner region is dominated by residual systematics and has very little disk flux. As a result, subtracting the models from the $Q_r$ image has little effect on this region, as demonstrated by the fact that the percentiles and the RMS in the residual images remain close to the original $Q_r$ values. The differences between the noise in $Q_r$ image and the $U_r$ images at these separations may be due to how the residual instrumental polarization affects each image. However, this remains to be confirmed. In the outer region, after the models have been subtracted all three quantities converge to the $U_r$ values, indicating that we are successfully subtracting off the disk flux with the residuals consistent with pure noise (i.e. the $U_r$ image).}


The best-fit HG (HG + Rayleigh) model reveals a disk inclination of $72\fdg2\pm0\fdg4$ ($68\fdg3^{+0\fdg7}_{-0\fdg8}$), with a relatively steep radial density power law index of $2.25\pm0.15$ ($2.2\pm0.2$), that extends from an inner radius of $81\pm2$ AU ($78.9\pm0.8$ AU) to an outer radius of $216\pm16$ AU ($211^{+21}_{-15}$ AU). The best-fit inclinations are roughly consistent with that determined by measuring the aspect ratio in the STIS disk image ($\sim75\degr$; Padgett et al., in prep). The outer radius is constrained by the level of forward scattering seen in the $Q_r$ image towards the NE, which quickly reaches the noise floor. Therefore, this fit parameter is governed by the sensitivity of these observations, and represents a lower limit on the true disk outer radius. Indeed, the disk is seen to extend to a radius greater than 500~AU in the HST/STIS images. However, it is possible that the HST/STIS observations probe a different dust population. 

\begin{deluxetable*}{lcccccc}

\tablecaption{Pixel Statistics for the data and model residuals images \label{tab:resids_stats}} 
\tablehead{
& \multicolumn{3}{c}{Inside 0\farcs275} & \multicolumn{3}{c}{Between 0\farcs275 and 1\farcs3} \\ \cline{2-4} \cline{5-7}\\
\colhead{Frame} & \colhead{5\%-tile} & \colhead{95\%-tile} & \colhead{RMS} & \colhead{5\%-tile} & \colhead{95\%-tile} & \colhead{RMS}}
\startdata
$Q_r$ & -26.0 & 18.3 & 13.6 & -3.1 & 8.6 & 3.8 \\
$U_r$ & -18.5 & 24.1 & 12.8 & -3.9 & 3.4 & 2.4 \\
HG Residuals & -27.4 & 14.4 & 13.4 & -4.1 & 3.3 & 2.4\\
HG+Rayleigh Residuals & -26.2 & 15.6 & 13.2 & -3.8 & 3.7 & 2.4\\
\enddata
\tablecomments{All values in this table are presented in raw data units (i.e. ADU/coadd).}
\end{deluxetable*}

The results of the disk fitting also indicate that the disk is offset from the star the along the projected semi-major axis, $\Delta X_1$, by $1.6\pm0.6$ AU ($1.4\pm0.6$ AU). This feature is consistent between both models, with both fits finding an offset with a significance slightly below $3\sigma$. The direction of this offset is consistent with the direction of the brightness asymmetries seen in Figure~\ref{fig:diskimg} and causes a faint brightness asymmetry in both model images in Figure~\ref{fig:resids}.

Visually, there are two major differences between the two models. First, there is an apparent deficit of light near the projected semi-minor axis in the HG + Rayleigh model which is due to the low polarization fraction at small (and large) scattering angles of the Rayleigh function. The deficit is in an area that coincides with a region of reduced S/N in Figure~\ref{fig:SNR} and it is therefore likely that this area was down-weighted in the MCMC fitting relative to the ansae. In Figure~\ref{fig:diskimg}, the brightness appears to increase towards the NE, with a maximum near the semi-minor axis and no apparent deficit. This may be a real feature of the disk, but could also be possibly due to residual systematics in the image. Indeed, there are strong residuals at similar separations along the SW semi-minor axis for both models.  Comparing the residuals between the two models in Figure~\ref{fig:resids}, there appears to be additional flux at the location of the deficit in the HG + Rayleigh model. However, the nearly identical reduced-$\chi^2$ value indicate that both model fit the data equally well. 

The second major difference between the two models is the brightness of the backside of the disk, which is barely visible in the HG + Rayleigh model but appears in the image of the HG model. A close examination of the HG + Rayleigh model residuals image, reveals a faint brightness at the location of the backside in the HG model that are not apparent in the HG-only model residuals. These residuals may indicate a very low S/N detection of the backside of the disk in the $Q_r$ image. 

Between the two models, the $\Delta X_2$ offset varies in both magnitude and sign. In the HG-only best-fit model $\Delta X_2$ is positive, indicating that the star is offset away from the observer along the inclination vector, relative to the disk center. The HG + Rayleigh model is best fit by a negative offset, where the star is offset towards the observer. This discrepancy illuminates a degeneracy between this offset and the exact form of the polarized scattering function. We therefore consider this parameter to remain poorly constrained. Future studies that are able to constrain the grain scattering properties or image the backside of the disk will be able to elucidate this remaining unknown. 

\section{Discussion}
\label{sec:discussion}
The dust seen in scattered light images of debris disks is thought to have originated in disks or belts of planetesimals, where collisional cascades grind km-sized bodies down to micron-sized dust \cite[see][and references therein]{Wyatt2008}. In order to initiate these cascades, the constituent planetesimals must be dynamically stirred such that their eccentricities reach a high enough level (on the order of $10^{-3}$ to $10^{-2}$) to allow their orbits to cross and for their collisions to be destructive. Stirring mechanisms include: self-stirring, where objects on the order of 2000~km located inside the belts induce the cascade from within \citep[e.g.][]{KenyonBromley2004}; planet-induced stirring, where a nearby planet excites the disk \citep[e.g.][]{Mustill2009}; or dynamical interaction with a passing star \citep[e.g.][]{KenyonBromley2002}. With the exception of the stellar fly-by scenario, the strength of all of these mechanisms should diminish with stellar age. As a given system reaches a steady-state configuration it cools dynamically and the collision rate slows. Therefore, the scattered light luminosity should dim with age as the small grains are removed via radiation pressure, Poynting--Robertson drag and/or stellar winds, and can no longer be replenished through collisions. Note that for disks of all ages, large collisions or other transient events may cause a a temporary increase in dust production and create a short-term increase in disk brightness \citep[e.g.][]{Wyatt2007}. 

The only published age estimates of HD~157587 are presented by \citet{Feltzing2001} and \citet{Casagrande2011}. \citet{Feltzing2001} fit  metallicity, effective temperature and absolute magnitude to the outputs of a rapid stellar evolution algorithm \citep{Hurley2000} based on the evolutionary tracks produced by \citet{Pols1998}. They estimate HD~157587 to have an age of $2.2\pm0.5$ Gyr. \citet{Casagrande2011} use a Bayesian analysis to fit effective temperature, metallicity and Johnson $V$ magnitude to the Padova evolutionary tracks \citep{Bertelli2008, Bertelli2009} and the BASTI isochrones \citep{Pietrinferni2004,Pietrinferni2006,Pietrinferni2009}. They find an age of 3.0$^{+1.7}_{-1.5}$\,Gyr and 3.0$^{+1.0}_{-1.4}$\,Gyr at 95\% confidence for the Padova and BASTI models, respectively.

Such advanced ages suggest that the star has evolved along the HR diagram away from the zero age main sequence (ZAMS). Alternatively, this offset from the ZAMS may be indicative of a pre-main sequence star moving towards the ZAMS, rather than away from it. However, considering the timescales of evolution along the pre-main sequence tracks ($\sim12$ Myr for an F5 star, \citealt{Siess2000}) compared to main sequence tracks (on the order Gyrs), it is more likely that the star has been found as it moves away from the ZAMS. 

Conversely, applying the star's proper motion, radial velocity and parallax to the BANYAN II webtool \citep{Gagne2014, Malo2013}\footnote{http://www.astro.umontreal.ca/~gagne/banyanII.php} indicates a 91\% probability that the star is a young ($< 1$ Gyr) field star, and a 9\% probability that the star is an old field star. Additionally, with velocities of [$U,V,W$] = [$-7, -17, -8$] km/s \citep{Holmberg2009}, we find that HD 157587 could be kinematically associated with several relatively young moving groups.  For example, in \citet{Chereul1999} the Pleiades stream 2-5 has [$U,V,W$] = [$-12.0\pm5.3, -21.6\pm4.7, -5.3\pm5.9$] km/s, whereas the Centaurus-Lupus stream 2-12 has [$U,V,W$] = [$-12.4\pm6.1, -16.5\pm4.6, -7.4\pm3.1$] km/s. In \citet{Asiain1999}, the Pleiades moving group B2 has  [$U,V,W$] = [$-10.7\pm5.3, -18.8\pm3.7, -5.6\pm2.2$] km/s, whereas members of Lower Centaurus Crux have [$U,V,W$] = [$-6.8\pm4.7, -18.5\pm6.5, -6.4\pm1.7$] km/s.  In all of these cases the age distributions determined by various methods lie in the range $10^7 - 10^9$ years. Thus, the stellar kinematics possibly support a relatively young age for HD~157587, while its photometry supports a much older age.

If HD~157587 is truly $2.5-3$ Gyr old, then its disk's optical/NIR emission would be unusually bright when compared to the overall population of imaged debris disks \citep[for a good summary see Figure 1 from][]{Choquet2016}. In fact, the disk around HD~157587 would represent one of only five debris disks imaged in scattered light with ages greater than one Gyr. The other four old disks seen in scattered light are HD~207129 \citep{Krist2010}, HD~202629 \citep{Krist2012}, HD~53143 \citep{Kalas2006} and HD~10647 \citep[q$^1$ Eri;][]{Stapelfeldt2007}. With a spectral type of F5V, HD~157587 is the earliest type star of the five, making it potentially the earliest type star older than one Gyr with a debris disk seen in scattered light. 

Of the other four old scattered-light debris disks (i.e. $> 1$ Gyr), HD~10647 is the most similar to HD~157587. HD~10647 is an F8V star with a debris disk at a radial distance of $\sim$85 AU \citep{Liseau2010}. The disk has been imaged in scattered light with HST/ACS \citep{Stapelfeldt2007} and in the infrared with Herschel/PACS \citep{Liseau2010}. As seen here for HD~157587, the HST images show evidence of a brightness asymmetry between the two ansae of the disk. Interestingly, HD~10647 is known to host a Jupiter mass planet at 2~AU. However, this planet is at too great a distance from the disk to have significant dynamical influence on the disk. \citet{Liseau2008} posit that the disk asymmetry suggests the presence of a second planet at larger distances. 

In our images of the HD 157587 disk we see a slight brightness asymmetry between the SE and NW ansae of the disk that we have modeled as being caused by a stellocentric offset. If the offset is due to a perturbations from a substellar companion, it is highly unlikely that it is one of the three point sources imaged in total intensity. Using the COND evolutionary models \citep{Baraffe2003} and assuming an age of 3 Gyr, we find their luminosities (flux ratios) correspond to brown dwarf masses between 30 and 40 $M_{Jup}$. If these three objects reside within the plane of the disk, their deprojected separations correspond to stellocentric radii between 210 AU and 380 AU, assuming circular orbits. Considering the smooth radial extent of the disk seen in the STIS images, we deem such an alignment unlikely, as the presence of such massive companions would cause significant disturbances to the disk morphology. Though it is possible that the expulsion timescale for the STIS grains are short enough to hide any such feature. 

Future imaging observations that reveal the relative motion of these three sources are required to understand their true relationship to HD~157587, if any. However, it is worth noting that the likelihood of a field star flyby is rare - less than $1\%$ in 100 Myr of a 500 AU approach \citep{KenyonBromley2002} - and considering HD~157587's proximity to the galactic plane, these three objects are likely background sources. If HD~157587 is a younger star, as suggested by the stellar kinematics, it is possible that a lower mass perturber at smaller angular separations is still bright enough in thermal emission to be detected via direct imaging. 
 
\section{Conclusion}

Using GPI we have imaged the dust ring around HD~157587 in $H$-band polarized intensity. The image reveals an inclined disk that appears to be cleared of material inside of a projected major axis of $\sim 80$ AU. The FOV of our observations overlaps with the inner regions of previous STIS images of the disk, and our analysis returns a similar disk inclination to that derived with the STIS data. The disk has a strong polarized brightness asymmetry in the NE-SW direction, where we interpreted the bright side of the disk to be tilted towards the observer. A similar brightness asymmetry has been seen in polarized observations of a number of other recently imaged disks, suggestive of similar grain compositions, size distributions and/or dust grain morphologies. Future detailed studies of these disks' dust composition that include multicolour observations or polarization fraction measurements will be able to further explore the similarities and differences of their dust grain populations. 

A second, weaker, brightness asymmetry is seen between the two ansae that could be due to a stellocentric offset in the plane of the sky. To test this hypothesis we used Bayesian MCMC methods to fit the polarized disk image to two disk models, one that used a HG polarized scattering phase function and one that combined a HG function with Rayleigh scattering phase function. Both models reveal an offset dust disk with an inner radius of 80 AU and an inclination of about $70\degr$. The center of the disk is found to be offset approximately 1.5 AU from the star's location in the plane of the sky and both models reproduce the brightness asymmetry between the two ansae. This offset could be confirmed with longer wavelength imaging using ALMA, which would trace thermal emission and therefore have less of a dependence on the scattering properties of the grains. 

In general the two model fits return similar disk properties, with the exception of $\Delta X_2$, the offset in the disk plane. We find that the form of the polarized scattered phase function is degenerate with the magnitude and direction of this offset and without further information on the form of the scattering phase function this value will remain poorly constrained.

The total intensity observations are dominated by stellar residuals at the location of the disk in the polarized intensity image and no disk was recovered. However, three point sources were recovered. Considering HD~157587's proximity to the galactic plane and the positions of the point sources relative to the disk, we consider these point sources to be background objects. Nonetheless, follow up observations are required to confirm this proposition. 

The currently published ages of the system that rely on stellar evolutionary tracts indicate an age well over 1 Gyr old. However, such an evolved age is at odds with the stellar kinematics. The stellocentric offset, suggest that this system has a complicated dynamical history and may harbour one or more unseen planets. This notion is reinforced by the similarities between HD~157587's stellar properties and disk morphology, and those of the RV planet host HD~10647. If the stellocentric offset is due to perturbations by one or more planets, further detailed study of the system's debris disk will be required to thoroughly characterize the system; the system's advanced age would make it ill-suited for direct imaging planet searches and radial velocity measurements would require prohibitively long time baselines. On the other hand, if the disk is younger than 1 Gyr, as implied by the stellar kinematics, then it presents itself as a prime target for deeper direct imaging observations which may be able to image the disk's perturber. 

\acknowledgments
This work was based on observations obtained at the Gemini Observatory, which is operated by the Association of Universities for Research in Astronomy, Inc., under a cooperative agreement with the National Science Foundation (NSF) on behalf of the Gemini partnership: the NSF (United States), the National Research Council (Canada), CONICYT (Chile), the Australian Research Council (Australia), Minist\'{e}rio da Ci\^{e}ncia, Tecnologia e Inova\c{c}\~{a}o (Brazil) and Ministerio de Ciencia, Tecnolog\'{i}a e Innovaci\'{o}n Productiva (Argentina).

This work was supported in part by NASA's NEXSS program, grant number NNX15AD95G. Portions of this work were performed under the auspices of the U.S. Department of Energy by Lawrence Livermore National Laboratory under Contract DE-AC52-07NA27344. Portions of this were were also carried out with the support of NSF grants AST-1413718 and AST-1518332, and NASA grant NNX15AC89G. This research used resources of the National Energy Research Scientific Computing Center, a DOE Office of Science User Facility supported by the Office of Science of the U.S. Department of Energy under Contract No. DE-AC02-05CH11231.

\facility{Gemini:South(GPI)}

\bibliographystyle{apj}   
\bibliography{references} 

\appendix


\begin{figure*}[!htbp]
\includegraphics[width=\linewidth]{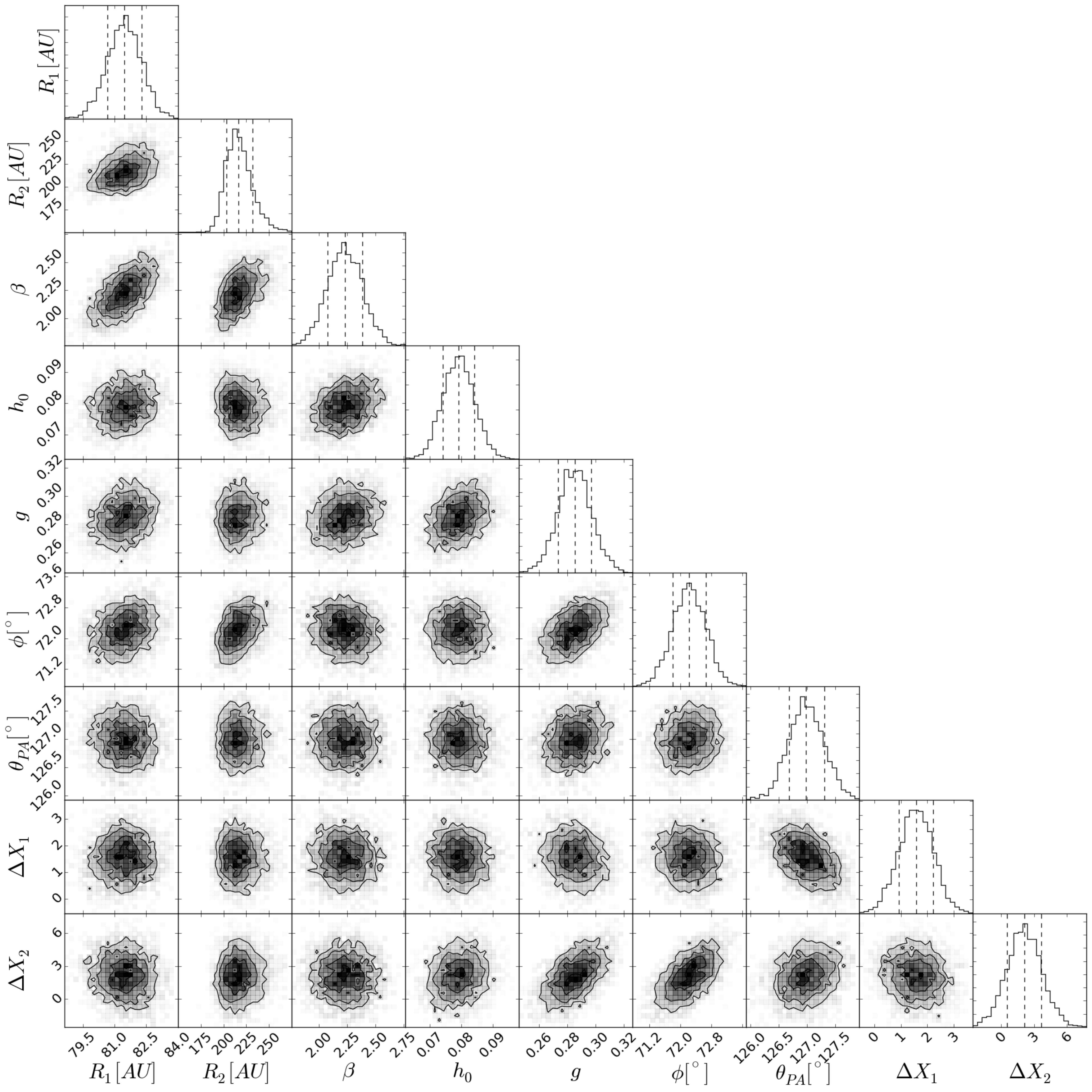}
\caption{The posterior distributions of the model parameters from MCMC disk model fitting to the $Q_r$ disk image with the HG only model. The diagonal histograms show the posterior distributions of each parameter marginalized across all the other parameters. In each plot, the dashed lines indicate the 16\%, 50\% and 84\% percentiles. The off-diagonal plots display the joint probability distributions with contour levels at the same percentiles. The normalization term, $N_0$, has been excluded from this plot and is considered a nuisance parameter. \label{fig:diskfit_pdfs_HG}}
\end{figure*}

\begin{figure*}[!htbp]
\includegraphics[width=\linewidth]{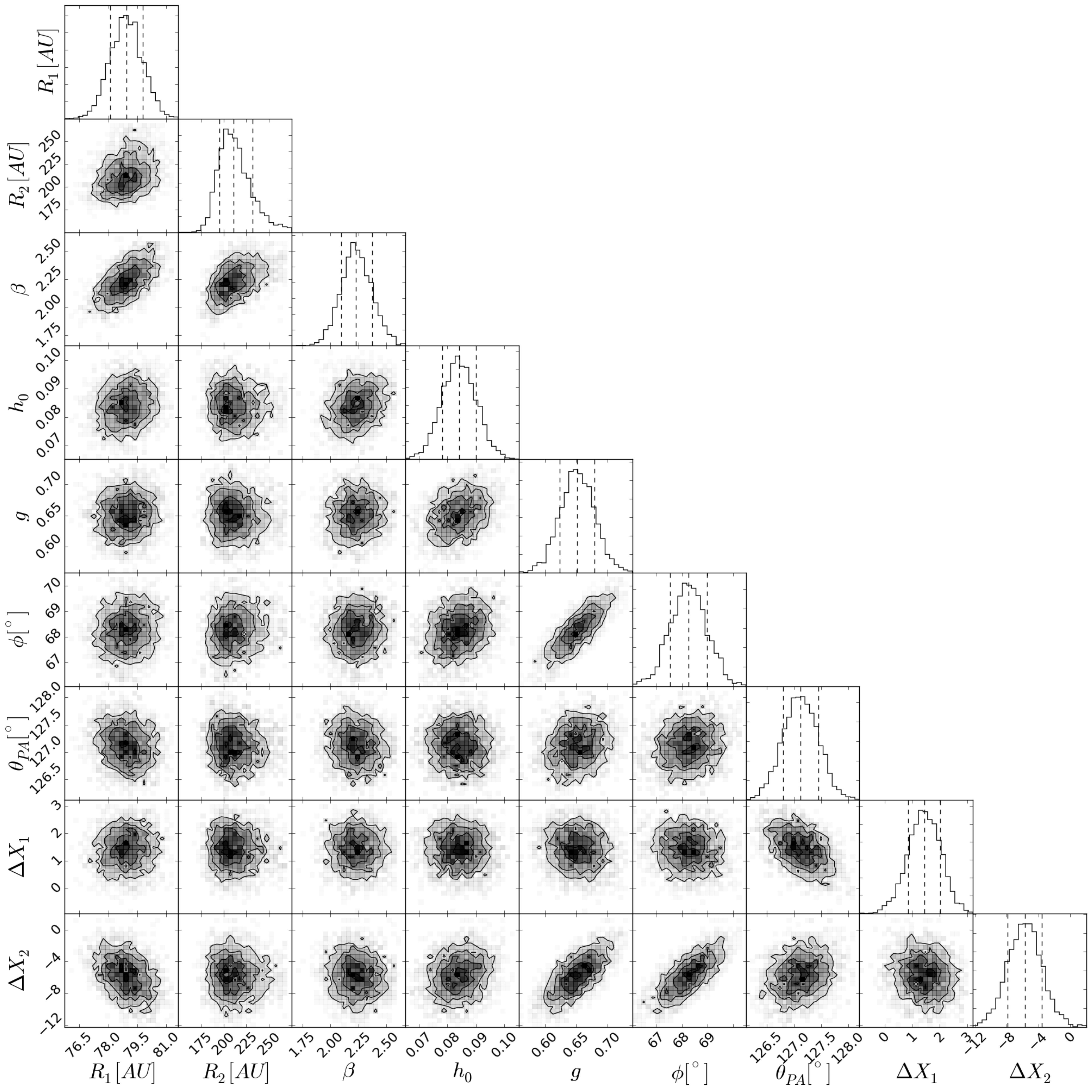}
\caption{The posterior distributions of the model parameters from MCMC disk model fitting to the $Q_r$ disk image with the HG + Rayleigh model. The diagonal histograms show the posterior distributions of each parameter marginalized across all the other parameters. In each plot, the dashed lines indicate the 16\%, 50\% and 84\% percentiles. The off-diagonal plots display the joint probability distributions with contour levels at the same percentiles. The normalization term, $N_0$, has been excluded from this plot and is considered a nuisance parameter. \label{fig:diskfit_pdfs_HGRayleigh}}
\end{figure*}

\end{document}